\documentclass[prl,aps,notitlepage,nofootinbib,floatfix,twocolumn,superscriptaddress,twoside]{revtex4-1}

\pdfoutput=1
\usepackage{natbib}
\usepackage{amssymb}
\usepackage{amsmath}    
\usepackage{graphicx}   
\usepackage{verbatim}   
\usepackage{color}      
\usepackage{hyperref}   

\usepackage{array}
\usepackage{enumerate}
\usepackage{multirow}

 \usepackage[normalem]{ulem}
 \usepackage{mathtools}

\newcommand{\aap}{{Astron.~Astrophys.}}

\newcommand{\mnras}{{Mon.~Not.~R.~Astron.~Soc.}}
\newcommand{\jcap}{{J.~Cosmol.~Astropart.~Phys.}}

\begin{document}

\title{Matter bispectrum of large-scale structure with Gaussian and non-Gaussian initial conditions: Halo models, perturbation theory, and a three-shape model}
\author{Andrei Lazanu}
\email{Andrei.Lazanu@pd.infn.it}
\affiliation{Centre for Theoretical Cosmology, DAMTP, University of Cambridge, CB3 0WA, United Kingdom}
\affiliation{INFN, Sezione di Padova, Via Marzolo 8, I-35131 Padova, Italy}
\author{Tommaso Giannantonio}
\affiliation{Kavli Institute for Cosmology, IoA, University of Cambridge, Madingley Road, Cambridge CB3 0HA, United Kingdom}
\affiliation{Centre for Theoretical Cosmology, DAMTP, University of Cambridge, CB3 0WA, United Kingdom}
\affiliation{Universit\"{a}ts-Sternwarte, Ludwig-Maximilians-Universit\"{a}t M\"{u}nchen, Scheinerstra{\ss}e 1, D-81679 M\"{u}nchen, Germany}
\author{Marcel Schmittfull}
\affiliation{Berkeley Center for Cosmological Physics, Department of Physics and Lawrence Berkeley National Laboratory, University of California, Berkeley, California 94720, USA}
\affiliation{Institute for Advanced Study, Einstein Drive, Princeton, New Jersey 08540, USA}
\author{E.P.S. Shellard}
\affiliation{Centre for Theoretical Cosmology, DAMTP, University of Cambridge, CB3 0WA, United Kingdom}

\date{\today}

\begin{abstract}
We study the matter bispectrum of large-scale structure by comparing the predictions of different perturbative and phenomenological models with the full three-dimensional bispectrum from $N$-body simulations estimated using modal methods. We show that among the perturbative approaches, effective field theory succeeds in extending the range of validity furthest 
on intermediate scales, at the cost of free additional parameters.
 By studying the halo model, we show that although it is satisfactory in the deeply non-linear regime, it predicts a deficit of power on intermediate scales, worsening at redshifts $z>0$. By comparison with the $N$-body bispectrum on those scales, we show that there is a significant squeezed component underestimated in the halo model. On the basis of these results, we propose a new ``three-shape'' model, based on the tree-level, squeezed and constant bispectrum shapes we identified in the halo model; after calibration this fits the simulations on all scales and redshifts of interest. We extend this model further to primordial non-Gaussianity of the local and equilateral types by showing that the same shapes can be used to describe the additional non-Gaussian component in the matter bispectrum. This method provides a \textsc{Halofit}-like prototype of the bispectrum that could be used to describe and test parameter dependencies and should be relevant for the bispectrum of weak gravitational lensing and wider applications. 
\end{abstract}

\maketitle

\section{I. Introduction}
To date, power spectrum estimation has been the primary route through which to infer cosmological information from surveys of large-scale structure (LSS) and the cosmic microwave sky.   However, the matter and galaxy bispectra (the Fourier transforms of the three-point correlators) will play an increasingly important role in the science exploitation of new data surveys.  Higher-order correlators offer complementary information that breaks cosmological parameter degeneracies and probes primordial non-Gaussianity, as well as being a key diagnostic tool for nonlinear density distributions.   Recently, in Ref.~\cite{bislong} we have undertaken a comprehensive survey of perturbative and non-perturbative theoretical approaches to the matter bispectrum, presenting direct three-dimensional comparisons with the complete bispectrum obtained from $N$-body simulations~\cite{Schmittfull2013}.   Despite the apparent complexity of three-point correlator calculations, this work indicates that the bispectrum analysis is tractable for large-scale structure both empirically and theoretically; on the one hand, optimal and efficient modal methods can be used for bispectrum estimation, and on the other, we demonstrate that significant mathematical simplifications are available to characterise the nonlinear gravitational bispectrum. We extend our previous work \cite{bislong} to include the effects of primordial non-Gaussianity of the local and equilateral types, showing that the same simple structure is present in both Gaussian and non-Gaussian bispectra.  Already for the cosmic microwave background (CMB), the bispectrum has become an important quantitative tool, most notably from the \textit{Planck} satellite \cite{2015arXiv150201582P}, and observational results are starting to emerge for the galaxy bispectrum, such as a recent BOSS analysis \cite{GilMarin2015a}.
A detailed analysis and understanding of the matter bispectrum represents an important step forward towards a quantitative modelling of the galaxy bispectrum. Together with an improved formalism for the galaxy bias, this would provide a new way to disentangle degeneracies between cosmological parameters and to obtain more stringent constraints on primordial non-Gaussianity than those provided by the CMB.
Moreover, even without the additional inclusion of galaxy bias, an accurate model of the matter bispectrum would be useful to describe the bispectrum of weak gravitational lensing.

\section{II. Bispectrum definitions}
\begin{figure}
\begin{center}
\includegraphics[width=3.2in]{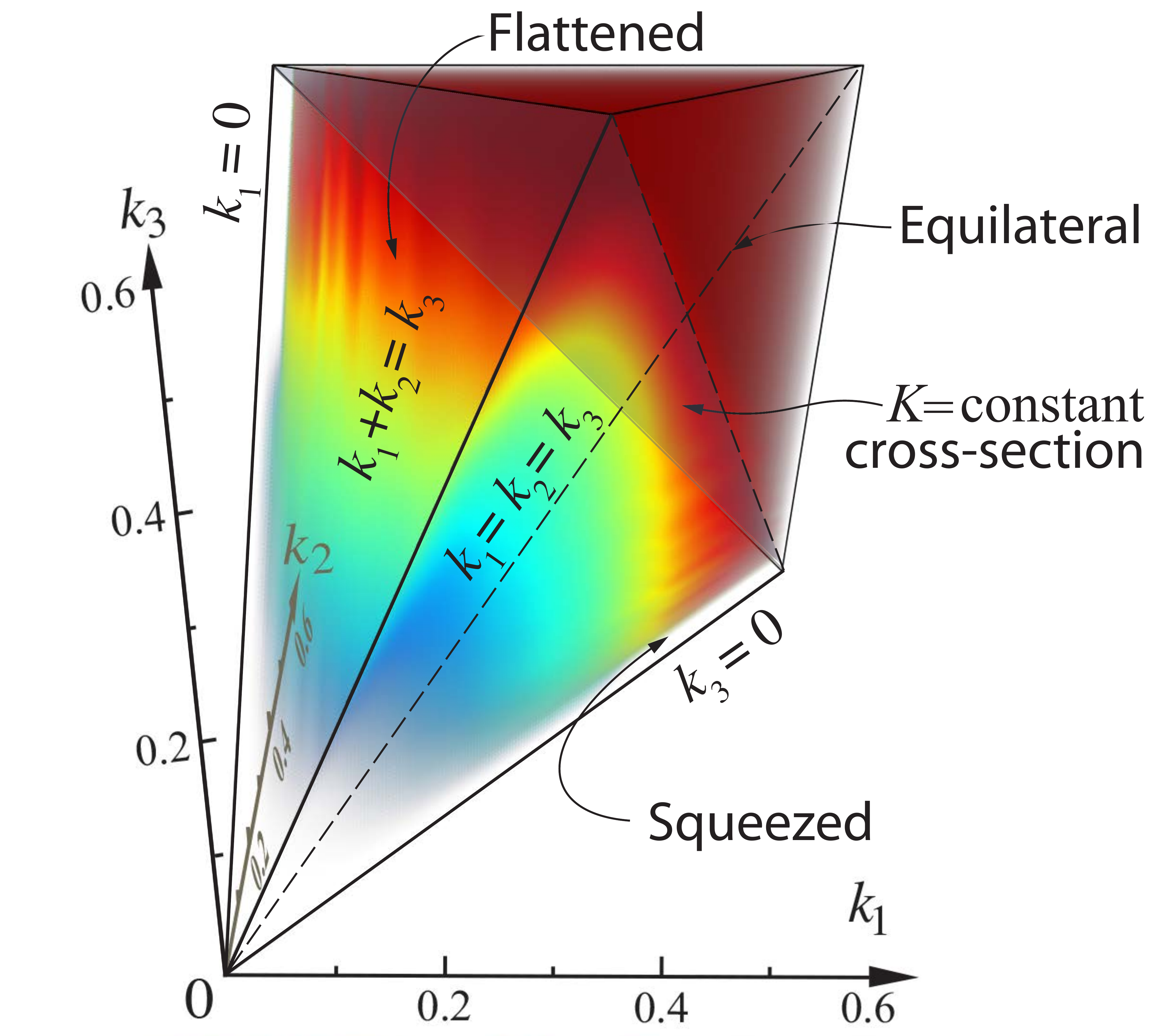}\\
\includegraphics[width=2.5in]{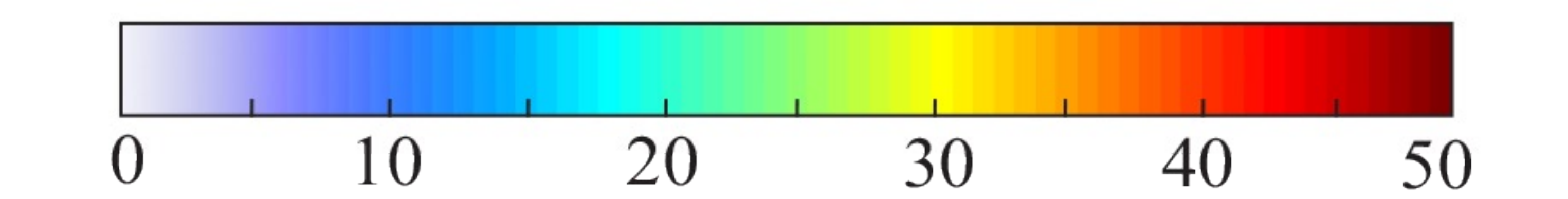}\\
\caption{The SN-weighted matter bispectrum from Gaussian $N$-body simulations at $z$$\,=\,$$0$, showing half the tetrahedral region on which it is defined (the front half of the bispectrum has been excised to show the interior signal).  The squeezed signal is located on the edges of the tetrapyd, the flattened signal on the faces, and the equilateral signal along $k_1$$\,=\,k_2$$\,=\,$$k_3$; this diagonal line is orthogonal to the $K\equiv k_1+k_2+k_3 = \hbox{const.}$ slicing. }
\label{bispectrum_data_0}
\end{center}
\end{figure}

The statistical analysis of random fields, such as the matter density perturbation $\delta \equiv \left(\rho - \bar \rho \right) / \bar \rho $, where $\rho$ is the matter density with mean $\bar \rho$, involves measuring its $N$-point correlation functions in real space or the equivalent polyspectra in Fourier space. We consider here the power spectrum and bispectrum, 
\begin{align}
\label{psbis}
\langle	\delta (\textbf{k}_1) \delta (\textbf{k}_2) \rangle &= (2\pi)^3 \delta_D (\textbf{k}_1 + \textbf{k}_2) P(k) \\
\langle	\delta (\textbf{k}_1) \delta (\textbf{k}_2) \delta (\textbf{k}_3) \rangle &= (2 \pi)^3 \delta_D (\textbf{k}_1 + \textbf{k}_2 + \textbf{k}_3) B(k_1,k_2,k_3) \, ,\nonumber
\end{align}
where $\delta_{D}$ is the Dirac delta function and  $\bf k_i$ are wavevectors with wavenumbers $k_i\,$$=$$\,|{\bf k}_i|$. For homogeneous and isotropic primordial perturbation models, to which we restrict ourselves, the bispectrum only depends on $k_1,k_2,k_3$. While the power spectrum is a function only of  $k$, the bispectrum is a three-dimensional function defined on a tetrahedral region for which $k_1,k_2,k_3$ satisfy the triangle condition in Eq.~(\ref{psbis}); the gravitational bispectrum obtained from $N$-body simulations is plotted in Fig.~\ref{bispectrum_data_0} to illustrate this domain. In an observational context, for comparing observations with theoretical models, it is useful to use an estimator that sums over all the relevant multipoles using an optimal signal-to-noise (SN) weighting. This leads naturally to defining a SN weighted scalar product between two bispectra $B_i$ and $B_j$ \cite{Fergusson2012}:  
\begin{equation}
\label{shapeprod}
\langle B_i, B_j \rangle \equiv \frac{V}{\pi}\int_{\mathcal{V}_B}dV_k\, \frac{k_1k_2k_3 \,B_i(k_1,k_2,k_3)\,B_j(k_1,k_2,k_3)}{ P(k_1)P(k_2)P(k_3)} \ ,
\end{equation}
where the integration domain $\mathcal{V}_B$ is the tetrahedral region shown in Fig.~\ref{bispectrum_data_0}.   Using the scalar product of Eq.~(\ref{shapeprod}),  we can define a \textit {shape correlator} $\cal S$ and a relative 
\textit{amplitude correlator} ${\cal A}$, respectively, as
\begin{align}
\label{correlators}
&\mathcal{S}\left(B_i,B_j\right) \equiv {\langle B_i, B_j \rangle}/{\sqrt{\langle B_i, B_i \rangle \langle B_j, B_j \rangle}} \, ,\nonumber \\
&\mathcal{A} \left( B_i, B_j \right) \equiv  \sqrt{ {\langle B_i, B_i \rangle}/{\langle B_j, B_j \rangle}} \, .
\end{align}
Both the amplitude and shape correlators can also be defined as binned or {\it sliced} quantities, by restricting the integration domain to narrow triangular slices at some $K \equiv k_1+k_2+k_3$ -- that is, on slices orthogonal to the tetrahedron diagonal $k_1=k_2=k_3$ (see Fig.~\ref{bispectrum_data_0}). The two correlators defined above quantify the error between the theoretical models and the simulations. The relative magnitude of the difference between bispectrum $B_i$ and $B_j$, representing the error of the theoretical modelling, can be expressed as  
\begin{equation}
\frac{\langle B_i-B_j, B_i-B_j \rangle}{\langle B_j, B_j \rangle} = 1 + \left( \mathcal{A}^2  -2 \mathcal{A}  \mathcal{S} \right) \left( B_i, B_j \right)  \ .
\end{equation}

As we shall see later in this paper, we can obtain an accurate global description of the non-linear gravitational  bispectrum from a sum of three simple bispectrum shapes $S^i(k_1,k_2,k_3)$, all motivated by comparison with the halo model, provided that we have the flexibility to calibrate an overall scale-dependent amplitude $f(K)$.  The constant shape is simply given by 
\begin{equation}
S^{\text{const}}(k_1,k_2,k_3)=1\,(\text{Mpc}/h)^6.
\label{constantsh}
\end{equation}
Physically, this corresponds to the late-time matter collapse in three dimensions on nonlinear scales. Comparing to the halo model, it is represented by the one-halo term.

The second shape is ``squeezed''  and can be expressed in terms of the linear power spectrum $P_0(k)$ as
\begin{multline}
S^{\text{squeez}} \left(k_1,k_2,k_3\right) = \\ ~~{\textstyle{\frac{1}{3}}} \left[P_0(k_1)P_0(k_2)  + P_0(k_2)P_0(k_3)+P_0(k_3)P_0(k_1) \right] \, .
\label{squeezed}
\end{multline}
In the halo model, this shape corresponds to the two-halo term. It is, however, older than the constant shape, being generated by the collapse of filamentary structures in two dimensions.

The third shape is the leading-order bispectrum of LSS:
\begin{align}
\label{stree}
&S^{\text{tree}} (k_1,k_2,k_3)=2P_0(k_1)P_0(k_2)F_2^{\left(s\right)} (\textbf{k}_1,\textbf{k}_2) +\\ 
& 2P_0(k_2)P_0(k_3)F_2^{\left(s\right)} (\textbf{k}_2,\textbf{k}_3)+ 2P_0(k_3)P_0(k_1)F_2^{\left(s\right)} (\textbf{k}_3,\textbf{k}_1)\, ,\nonumber
\end{align}
where the standard gravitational kernel $F_2^{(s)}$ is discussed in Ref.~\cite{bislong}. This shape is the well-known tree-level bispectrum of Eulerian perturbation theory. It provides a very good approximation to the three-halo term.
The actual bispectrum is more closely approximated by the tree-level shape [Eq.~(\ref{stree})] if we replace $P$ with the nonlinear \textsc{Halofit} ~\citep{2003MNRAS.341.1311S, 2012ApJ...761..152T} $P_\text{NL}$ power spectrum ~\cite{Scoccimarro21082001}; we denote the improved shape by $S^{\text{treeNL}}$.
Putting this all together, we obtain the ``three-shape'' model~\cite{bislong}:
 \begin{align}
 \label{shapes}
&B_\text{3-shape}(k_1,k_2,k_3)= 
f_{1h}(K) \, S^{\text{const}}(k_1,k_2,k_3) \\
&+  f_{2h}(K) \, S^{\text{squeez}}(k_1,k_2,k_3) 
 +   f_{3h}(K) \, S^{\text{treeNL}}(k_1,k_2,k_3)  \, .\nonumber
\end{align}

\section{III. Primordial non-Gaussianity}
The density perturbations $\delta$ can be expressed in terms of the primordial potential $\Phi$ using the Poisson equation:
\begin{align}
\delta(\textbf{k},z)=\frac{2}{3} \frac{k^2 T(k)D(z)}{\Omega_m H_0^2} \Phi(\textbf{k}) \equiv M(k,z)\Phi(\textbf{k}) \, ,
\end{align}
where $T(k)$ is the linear transfer function and $D(z)$ is the growth rate. Then the linear power spectrum and initial bispectrum can be expressed as
\begin{align}
P_0(k)&=M^2(k)P_{\Phi}(k) \, ,\\
B_0(k_1,k_2,k_3)&=M(k_1)M(k_2)M(k_3)B_{\Phi}(k_1,k_2,k_3) \, ,
\end{align}
where $P_{\Phi}$ and $B_{\Phi}$ are the primordial potential power spectrum and bispectrum respectively.
For local and equilateral types of primordial non-Gaussianity, $B_{\Phi}$ takes the following forms:
\begin{align}
B_{\phi}^{\text{local}}&(k_1,k_2,k_3)=2f_{\text{NL}}^{\text{local}} \left[P_{\phi}(k_1)P_{\phi}(k_2)+\text{2 perms} \right] \\
B_{\phi}^{\text{eq}}&(k_1,k_2,k_3)=6f_{\text{NL}}^{\text{eq}} \left\{-[P_{\phi}(k_1)P_{\phi}(k_2)+\text{2 perms} ]\right. \nonumber \\
&-2[P_{\phi}(k_1)P_{\phi}(k_2)P_{\phi}(k_3)]^{2/3} \nonumber \\
&+[P_{\phi}^{1/3}(k_1)P_{\phi}^{2/3}(k_2)P_{\phi}(k_3)+\text{5 perms}]\left. \right\} \,.
\end{align} 

\section{IV. Perturbation theory}
Perturbative approaches expand the density contrast $\delta$ in a series of perturbations $\delta^{(n)}$, which are valid
 while $|\delta| \ll 1$ ~ \cite{Bernardeau20021}; this is true at  early times and on large scales.
In Eulerian standard perturbation theory (SPT), expansions in terms of powers of the variable $\delta$ are considered. The leading-order tree-level bispectrum labelled $B_{211}$ is given by Eq.~(\ref{stree}) ~\cite{Fry1984}. 
For the one-loop correction in SPT, four additional terms are added to this, labelled $B_{222}$, $B_{321}^{(I)}$, $B_{321}^{(II)}$ and $B_{411}$~\cite{Bernardeau20021}.  These terms only succeed in extending the range of validity of the tree-level [Eq.~(\ref{stree})] by a small amount and yield an excess of power on quasi-nonlinear scales (essentially because loop integrals require integration of momenta over an infinite range where the assumption that $|\delta| \ll 1$ is no longer valid).

The observation that even when the density fluctuations are no longer small the gravitational potential still remains small, together with this excess, has led to the development of the effective field theory of LSS (EFT) ~ \cite{1475-7516-2012-07-051, Carrasco2012}. In this theory, there are additional parameters that quantify the effects of non-perturbative small-scale physics on large scales through an effective stress-energy tensor.
 The power spectrum and bispectrum are modified at leading order by adding terms to each of them ~\cite{Angulo:2014tfa,baldauf}, which include parameters calibrated on simulations. However, the leading EFT bispectrum counter-term is calibrated using just the power spectrum.

Another method that improves the accuracy of the prediction is renormalised perturbation theory (RPT) \cite{PhysRevD.73.063519}; the SPT expansion  is reorganised and then resummed. The calculations simplify significantly with the \textsc{MPTbreeze} method \cite{Crocce2012}, which expresses the power spectrum and the bispectrum directly in terms of the positive-definite  SPT terms, but multiplied by a decaying exponential function. Alternatively, Lagrangian perturbation theory displaces particles from their Lagrangian initial positions \cite{Ehlers1997}. By considering the expansion of the mass density in these displacements and using a suitable resummation, the resummed Lagrangian perturbation theory (RLPT) is obtained ~\cite{Matsubara2008, Rampf2012}. 

In Ref.~\cite{bislong}, we survey all the above-mentioned methods and calculate the bispectrum (even up to some two-loop terms).
By considering the sliced shape correlators $\mathcal{S}$ of each theory bispectrum with the three canonical shapes [Eqs.~(\ref{constantsh})-(\ref{stree})], we show that all the perturbative approaches can be well approximated by the `flattened' tree-level shape (\ref{stree}) (see Fig.~6 in Ref.~\cite{bislong}).

\begin{figure*}
\begin{center}
\includegraphics[width=0.3\linewidth]{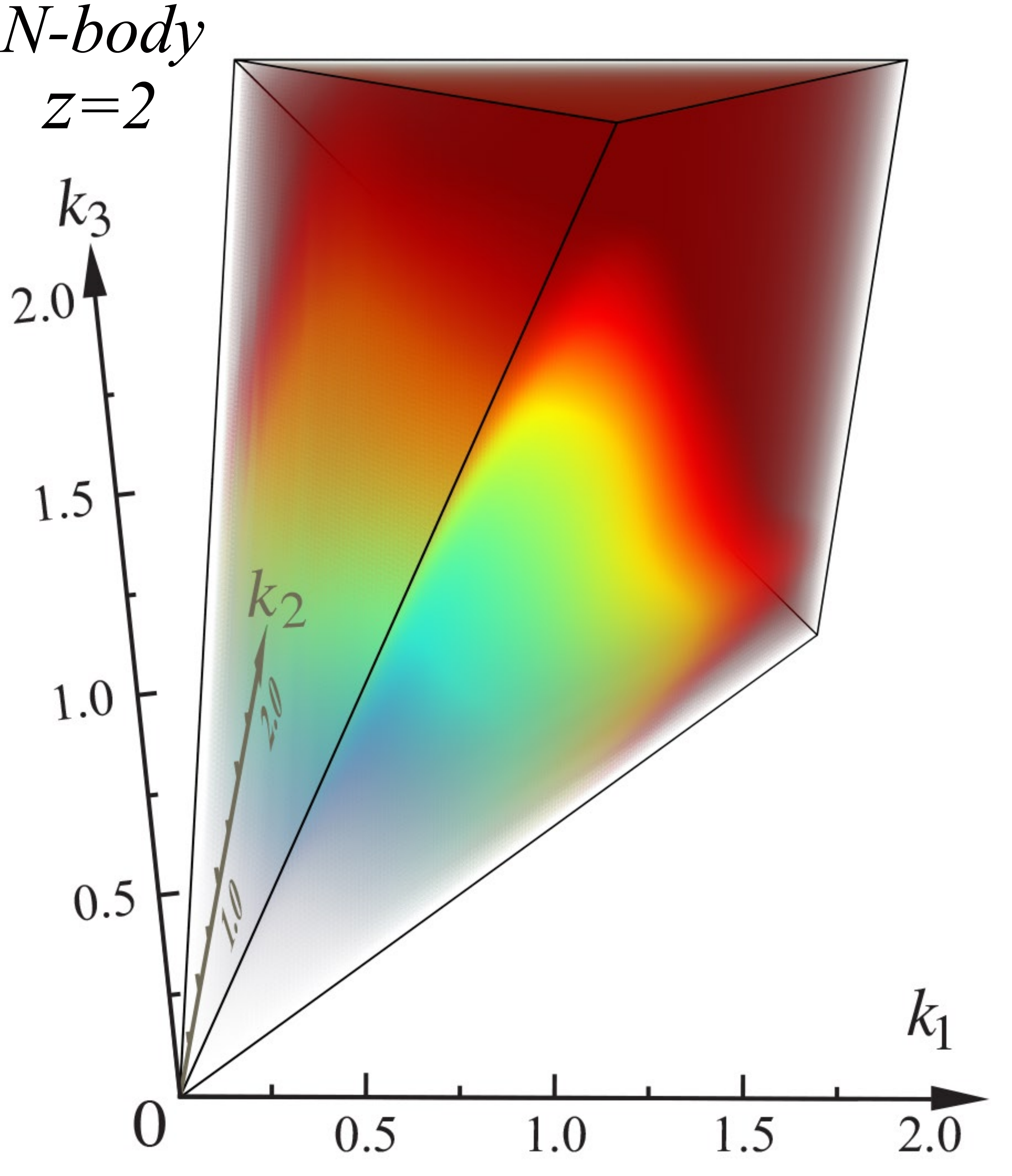} \includegraphics[width=0.3\linewidth]{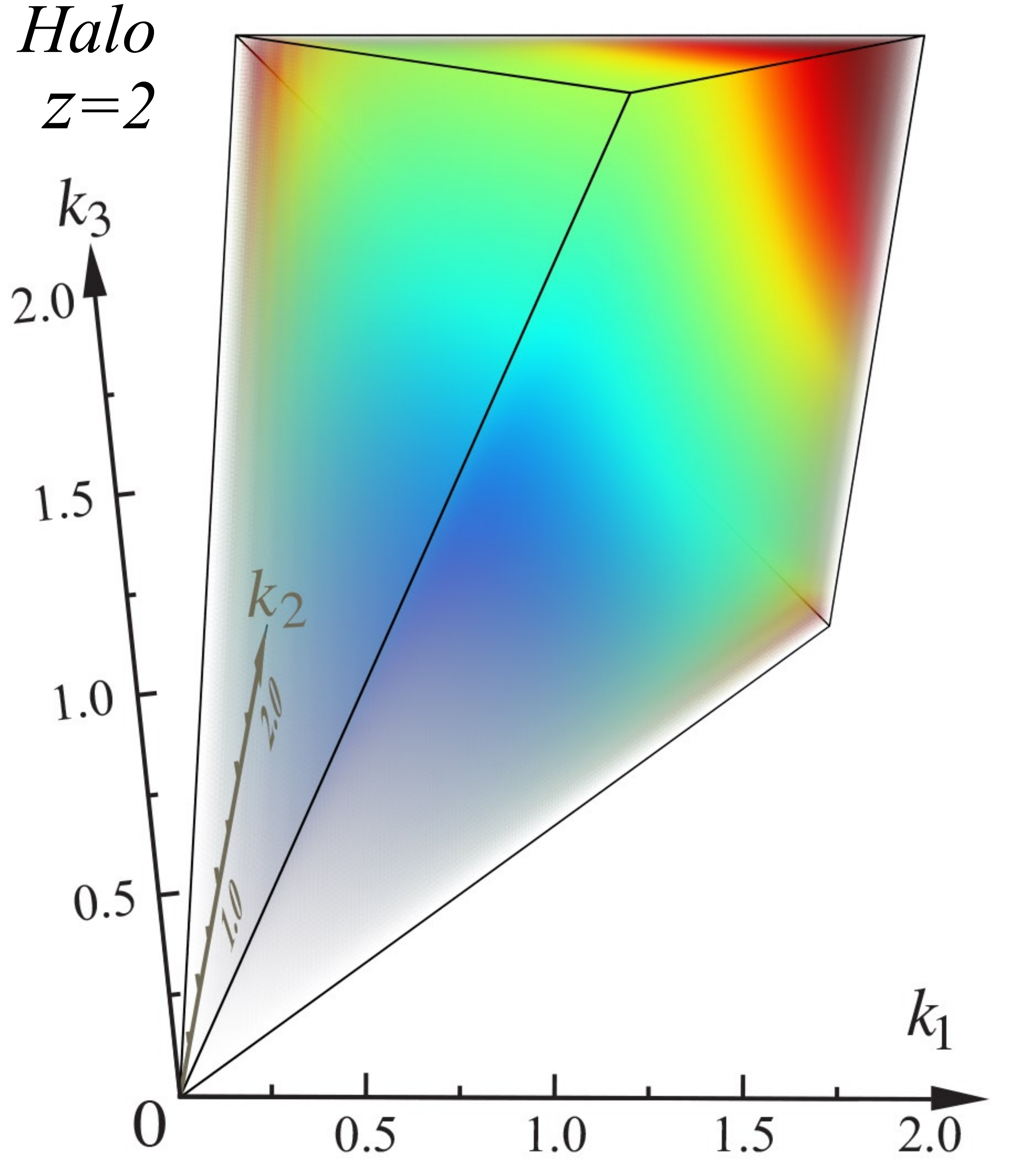} \includegraphics[width=0.3\linewidth]{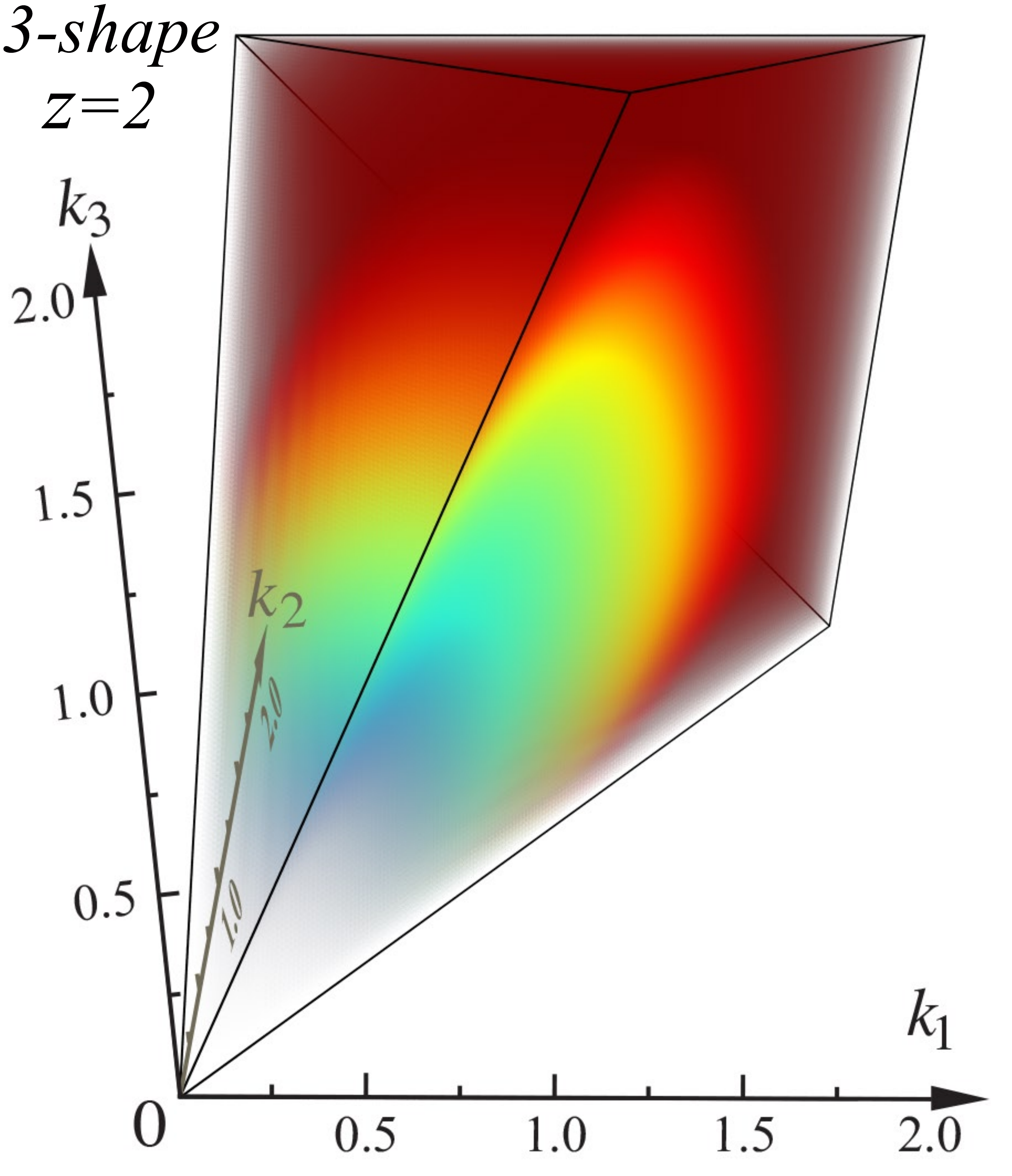} \\
\includegraphics[width=0.33\linewidth]{scale50.pdf}
\caption{The SN-weighted matter bispectrum at $z\,$$=$$\,2$ in the Gaussian case, $f_{\text{NL}}=0$: We compare the $N$-body bispectrum (left) with the standard halo model (middle) and the ``three-shape'' model (right).  The halo model for $z>0$ exhibits a large deficit (including a missing squeezed component), while the ``three-shape'' model shows improved agreement with the $N$-body results.}
\label{bispectrum_data2}
\end{center}
\end{figure*}

For primordial non-Gaussianity, we consider the corrections appearing at tree-level and at one loop in the SPT approach \cite{2010MNRAS.406.1014S}.
\section{V. Phenomenological models}
The halo model extends the modelling of dark matter clustering into the nonlinear regime by assuming that all dark matter in the Universe has collapsed into haloes ~\cite{Seljak11102000,Cooray20021}. The halo model power spectrum and bispectrum are composed of two and three terms, respectively, corresponding to pairs or triplets of particles residing
 in one, two, or three haloes. The three bispectrum components are labelled $B_{1h}$, $B_{2h}$ and $B_{3h}$, and they contribute mainly to the fully nonlinear regime, to intermediate scales and to the perturbative limit, respectively.

The halo model provides a satisfactory description of $N$-body simulations at $z\,$$=$$\,0$ in the fully nonlinear regime, but it has some well-known shortcomings even at late times~\cite{Cooray20021,  PhysRevD.78.023523}: (a) In the transition between the linear and non-linear regimes, the accuracy of the model is at the sub-$10\%$ level. (b) On very large scales $P_{1h}$, $B_{1h}$ and $B_{2h}$ do not decay to zero, thus leading to an excess of power with respect to linear theory.(c) The assumption that all matter in the Universe has collapsed into spherical haloes with negligible sub-structure at all redshifts leads to growing inaccuracies for $z>0$.

The first two problems can be improved by combining the halo model with perturbation theory ~\cite{valageas2}, which we call the ``halo-PT model''. This cuts off excess power in the one- and two-halo terms with a prescription that endeavours to ensure three particles contribute to only one of the 
terms. This modifies $B_{1h}$ and $B_{2h}$, while $B_{3h}$ is substituted with the perturbative prediction, for which we use  EFT in our implementation~\cite{bislong}.

By analysing the bispectrum shapes of the three halo model components, we find that the one-, two- and three-halo terms have excellent sliced shape correlations $\mathcal{S}$ [Eq.~(\ref{correlators})] with the constant [Eq.~(\ref{constantsh})], squeezed [Eq.~(\ref{squeezed})] and tree-level [Eq.~(\ref{stree})] shapes, respectively, for all redshifts considered (see Fig.~7 of Ref.~\cite{bislong}). 

The halo model can also be extended for the case of primordial non-Gaussianity, by a suitable modification of the halo profile, mass function and bias. The results presented in this paper are based on the methodology described in Ref. \cite{1475-7516-2012-08-036}.
\begin{figure}
\begin{center}
\includegraphics[width=0.83\linewidth]{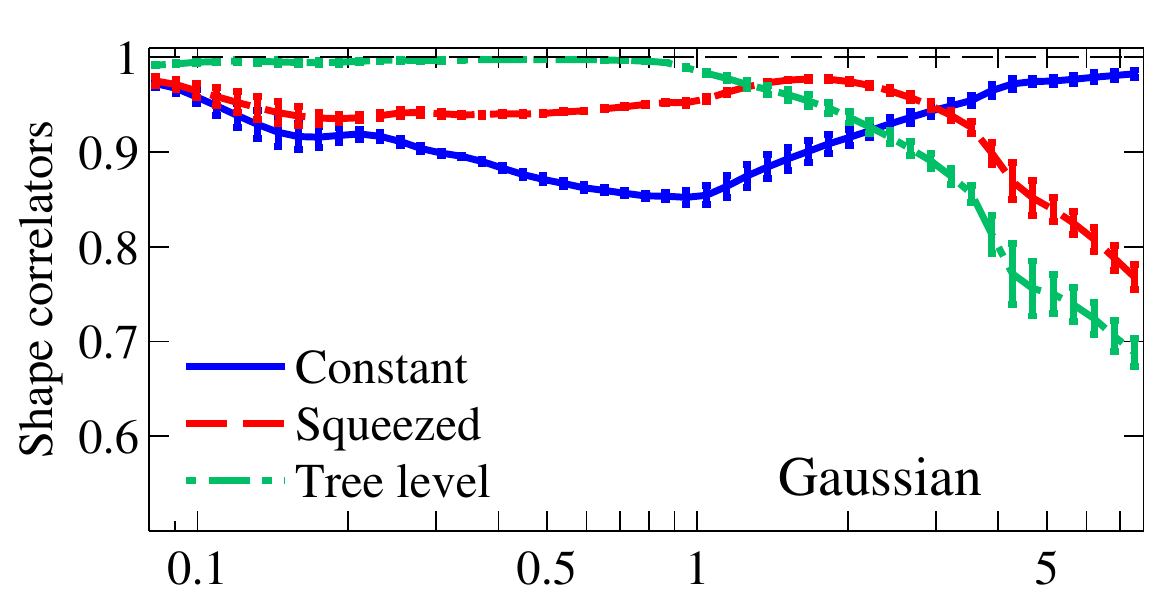}
\vspace{0cm}
\includegraphics[width=0.83\linewidth]{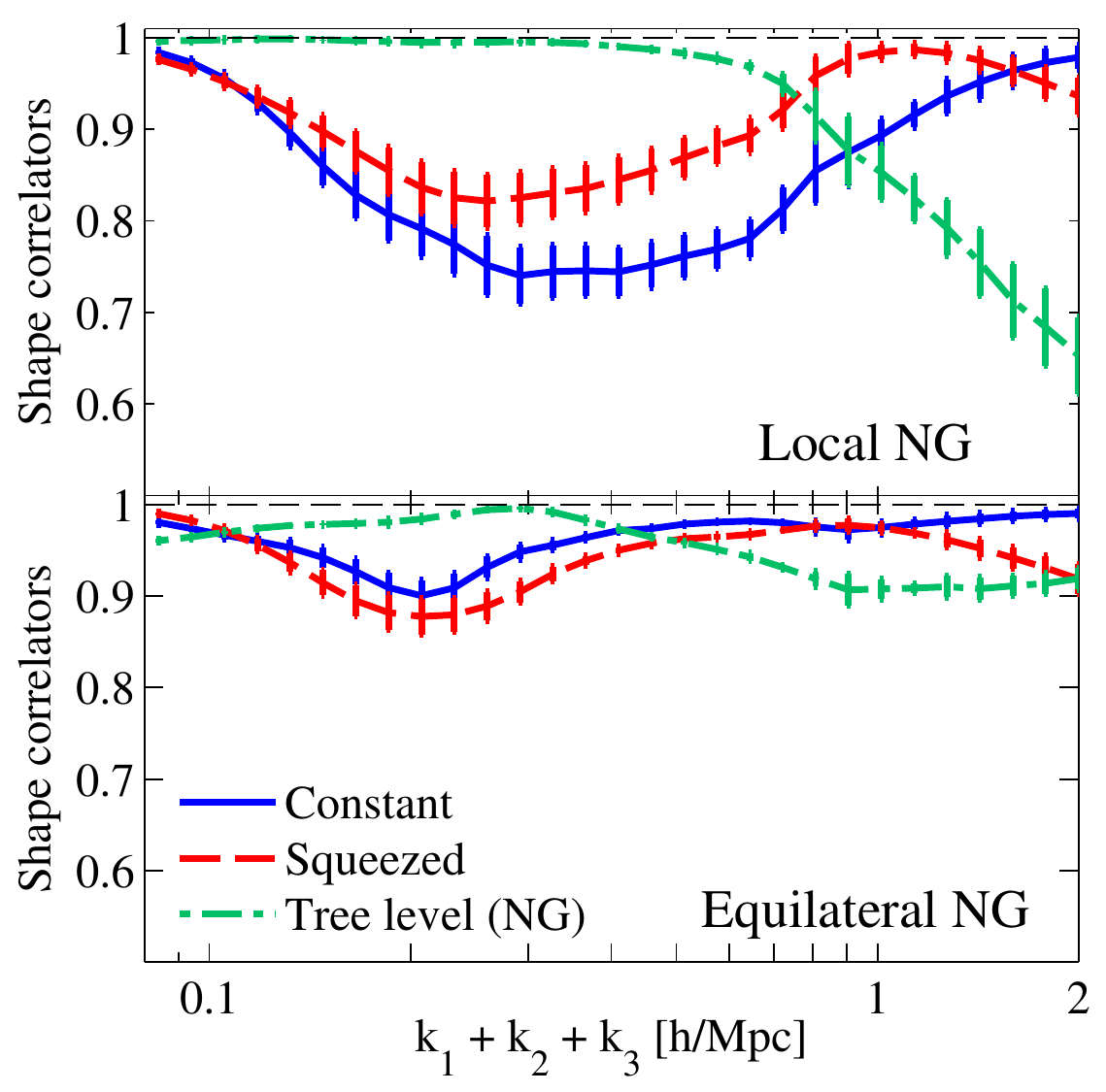}
\caption{Sliced shape correlations between the Gaussian and non-Gaussian corrections to $N$-body bispectra and the three canonical shapes [Eqs.~(\ref{constantsh})-(\ref{stree})] at $z=1$. A linear combination offers good shape correlations at all $k$ for the Gaussian (top), local (middle) and equilateral (bottom) non-Gaussian corrections. }
\label{bisshapez1}
\end{center}
\end{figure}

\begin{figure}
\begin{center}
\includegraphics[width=0.95\linewidth]{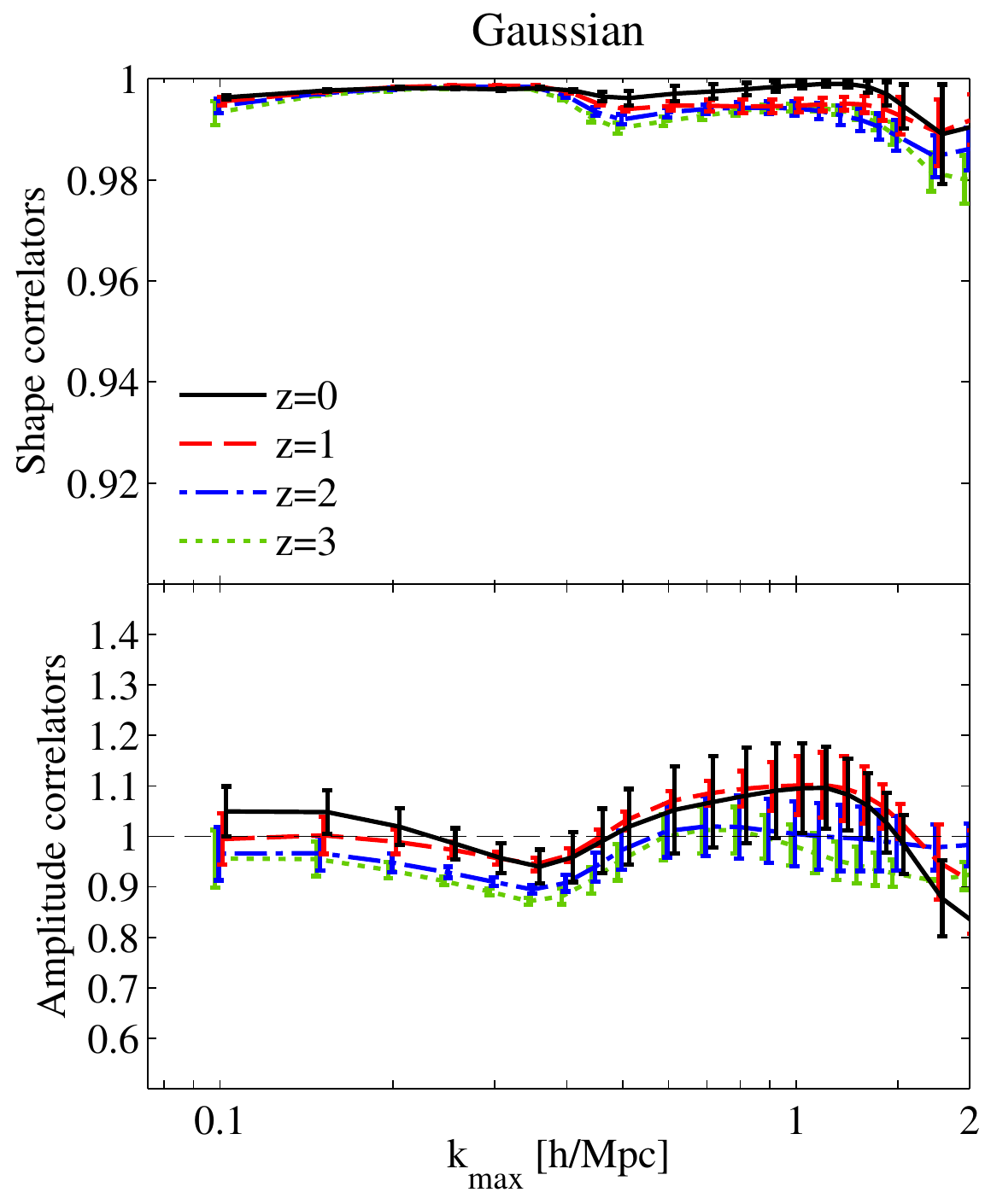}
\caption{Shape (upper) and amplitude (lower) correlators between the measured Gaussian $N$-body matter bispectrum and the ``three-shape'' model at four redshifts.   The correlators are defined with Eq.~(\ref{shapeprod}) integrated up to the wavenumber $k_{\max}$.}
\label{absb}
\end{center}
\end{figure}

\begin{figure}
\begin{center}
\includegraphics[width=0.95\linewidth]{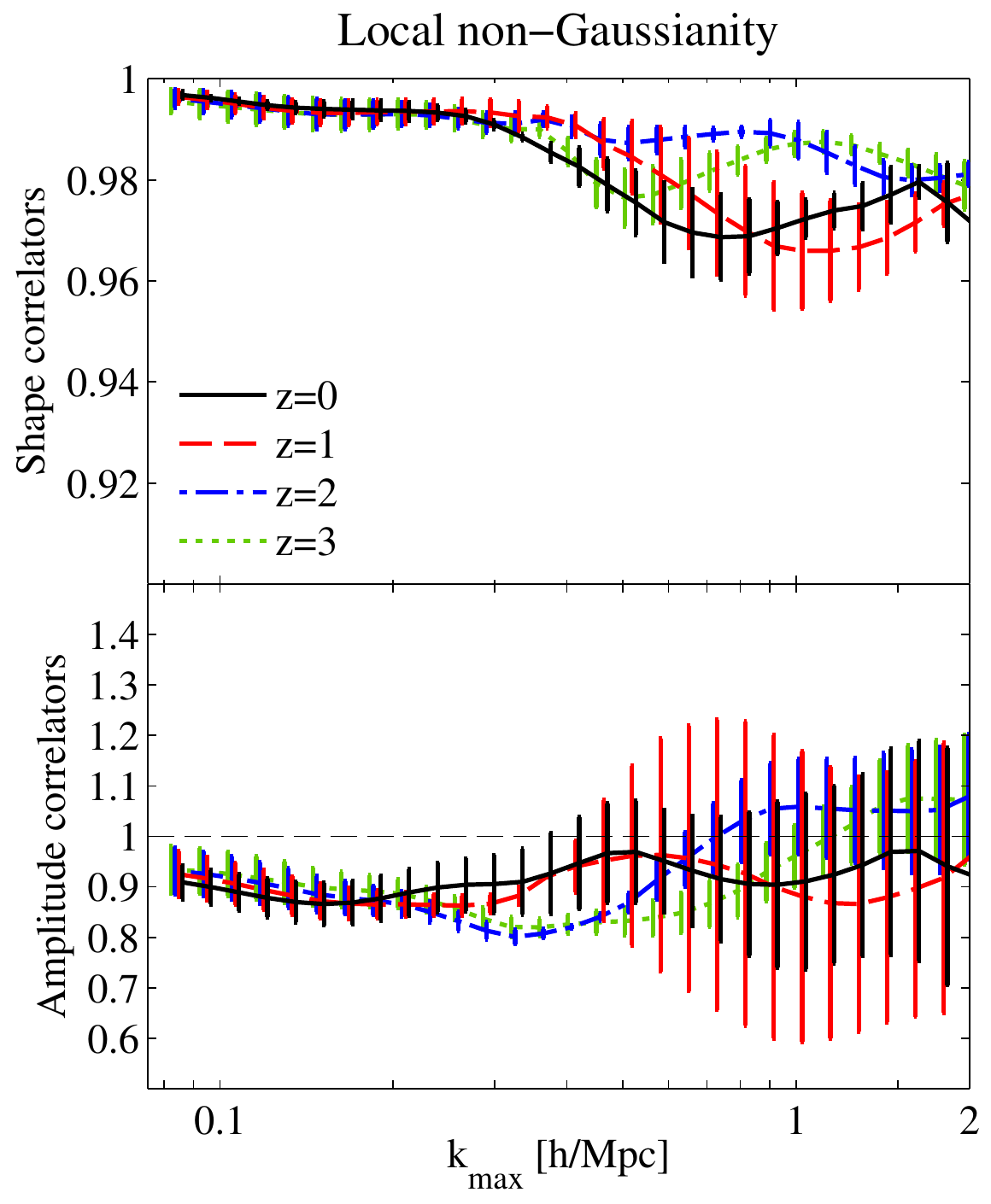}
\caption{Shape (upper) and amplitude (lower) correlators between the difference of the measured $N$-body matter bispectrum for $f_{\text{NL}}^{\text{local}}=10$ and the Gaussian simulations and the ``three-shape'' model [Eq. (\ref{shapesNG})] at four redshifts.   The correlators are defined with Eq.~(\ref{shapeprod}) integrated up to the wavenumber $k_{\max}$.}
\label{bisloc}
\end{center}
\end{figure}

\section{VI. Simulations and benchmark model}
To calculate the matter bispectrum, we analyse density distributions from three sets of $N$-body simulations with Gaussian and non-Gaussian initial conditions described in Ref.~\cite{Schmittfull2013}. For the non-Gaussian simulations, we use $f_{\text{NL}}^{\text{local}}=10$ and $f_{\text{NL}}^{\text{eq}}=100$.
We have used three realisations for each of our simulations, both for the Gaussian and for the non-Gaussian simulations. The error bars represent $1\sigma$ deviations from the mean of the realisations.
The signal-to-noise weighted bispectrum $\hat B$ is reconstructed using a separable modal basis $Q_n$~\cite{Fergusson2012} as 
\begin{equation}
\frac{\hat B (k_1,k_2,k_3) \, \sqrt{k_1k_2k_3}}{\sqrt{P(k_1) P(k_2) P(k_3)}} = \sum_{n=0}^{n_{\max}} \beta_n^Q \, Q_n(k_1,k_2,k_3) \, ,
\label{eigenfns}
\end{equation}
where about  $100$ $\beta_n^Q$ coefficients were required for good convergence. 
The three simulations cover the $k$-ranges $[0.0039, 0.5] \, h/\text{Mpc}$, $[0.016, 2.0] \, h/\text{Mpc}$ and $[0.062, 8.0] \, h/\text{Mpc}$, respectively, and are combined with a smoothing function, explained in Ref.~\cite{bislong}. 
Plots of the full 3D SN-weighted bispectrum obtained at $z=0,2$ in Figs.~\ref{bispectrum_data_0} and \ref{bispectrum_data2} (left) show qualitatively how shape changes with scale and redshift. 
The non-Gaussian correction is small compared to the Gaussian part of the signal, and hence it is not viable to calibrate the ``three-shape'' model directly. Instead, we choose to consider only the non-Gaussian part of the signal $\Delta B_{\text{NG}} \equiv B(f_{\text{NL}})-B(f_{\text{NL}}=0)$, and to determine a ``three-shape'' model for it. We choose the same shapes as for the Gaussian model for the constant (\ref{constantsh}) and squeezed shapes (\ref{squeezed}), but the tree-level shape (\ref{stree}) is replaced by $B_0$, denoted $S^{\text{tree,NG}}$ in analogy with the Gaussian result. Therefore, the ``three-shape'' model for primordial non-Gaussianity becomes
 \begin{align}
 \label{shapesNG}
&\Delta B_\text{3-shape}(k_1,k_2,k_3)= 
c_{1h}f_{1h}(K) \, S^{\text{const}}(k_1,k_2,k_3) \\
&+  c_{2h}f_{2h}(K) \, S^{\text{squeez}}(k_1,k_2,k_3) 
 +   S^{\text{treeNL,NG}}(k_1,k_2,k_3)  \, , \nonumber
\end{align}
where the non-Gaussian tree-level bispectrum has been boosted with the \textsc{Halofit} power spectrum, the functions $f_{1h}$ and $f_{2h}$ are the fits for the Gaussian simulations, and only two coefficients $c_{1h}$ and $c_{2h}$ require fitting to the simulated data.

A quantitative analysis of the $N$-body bispectrum for the Gaussian case using the sliced shape correlator [Eq.~(\ref{correlators})] is summarised in the top panel of Fig.~\ref{bisshapez1}. The flattened tree-level shape [Eq.~(\ref{stree})] is an excellent approximation at low wavenumbers $k$, while the constant shape [Eq.~(\ref{constantsh})] dominates for high $k$ on nonlinear scales. On intermediate scales, an additional squeezed component [Eq.~(\ref{squeezed})] prevails.  In combination, these results confirm the proposed ``three-shape'' model [Eq.~(\ref{shapes})], which we describe in more detail in Ref.~\cite{bislong}, together with the numerical values of the parameters used to calibrate simple forms of  $f_{1h}(K)$, $f_{2h}(K)$ and $f_{3h}(K)$.

On large scales, the ``three-shape'' model is based on the non-linear tree-level bispectrum (or any perturbative approach, e.g. EFT), but this leads to an excessive amplitude on intermediate scales.  To alleviate this problem requires a prescription to prevent double-counting in $f_{3h}$, so we introduce an exponential cut-off which adds one free parameter. On small scales, the one-halo bispectrum term provides an adequate description, so we use a simple fit $f_{1h}$ to this, accurate at all redshifts and introducing no extra parameters.
In the intermediate regime corresponding to the two-halo term, the standard halo model shows a power deficit that worsens at higher redshift (see Fig.~\ref{bispectrum_data2}, middle panel). We find that the two-halo deficit can be reduced by boosting this term by approximately $D(z)^{-1.7}$, where $D(z)$ is the linear growth factor normalised to unity today. However, this is not entirely satisfactory, because it increases the excess on small scales as $k \to 0$. 
To improve the accuracy of this term,  we model $f_{2h}$ with a simple function, scaling as $f_{2h} \propto K^3$ for $K \to 0$. We fix the functional form and introduce two free parameters calibrated to simulations. 
We show in Fig.~\ref{absb} that the ``three-shape'' model thus obtained fits the Gaussian simulations well for all wavenumbers $k$ and redshifts $z$ considered (shape correlation $\mathcal{S}> 98\%$ and relative amplitude  $\mathcal{A} \simeq 1$ within 10\%).

For primordial non-Gaussianity, a very similar shape description remains valid, the only significant difference being that for $\Delta B_{\text{NG}}$, one needs to replace the tree-level shape with a \textsc{Halofit}-enhanced tree-level non-Gaussian correction [Eq. (\ref{shapesNG})]. In Fig.~\ref{bisloc}, we show the fit for $\Delta B_{\text{NG}}^{\text{local}}(f_{\text{NL}}=10)$. The systematic offset appearing at $k_{\text{max}} \approx 0.3 \, h/$Mpc at higher redshifts is due to the less than perfect choice of functional form of $f_{2h}$ at these redshifts. A similar result is obtained for the equilateral simulation.

\begin{figure}
\begin{center}
\includegraphics[width=\linewidth]{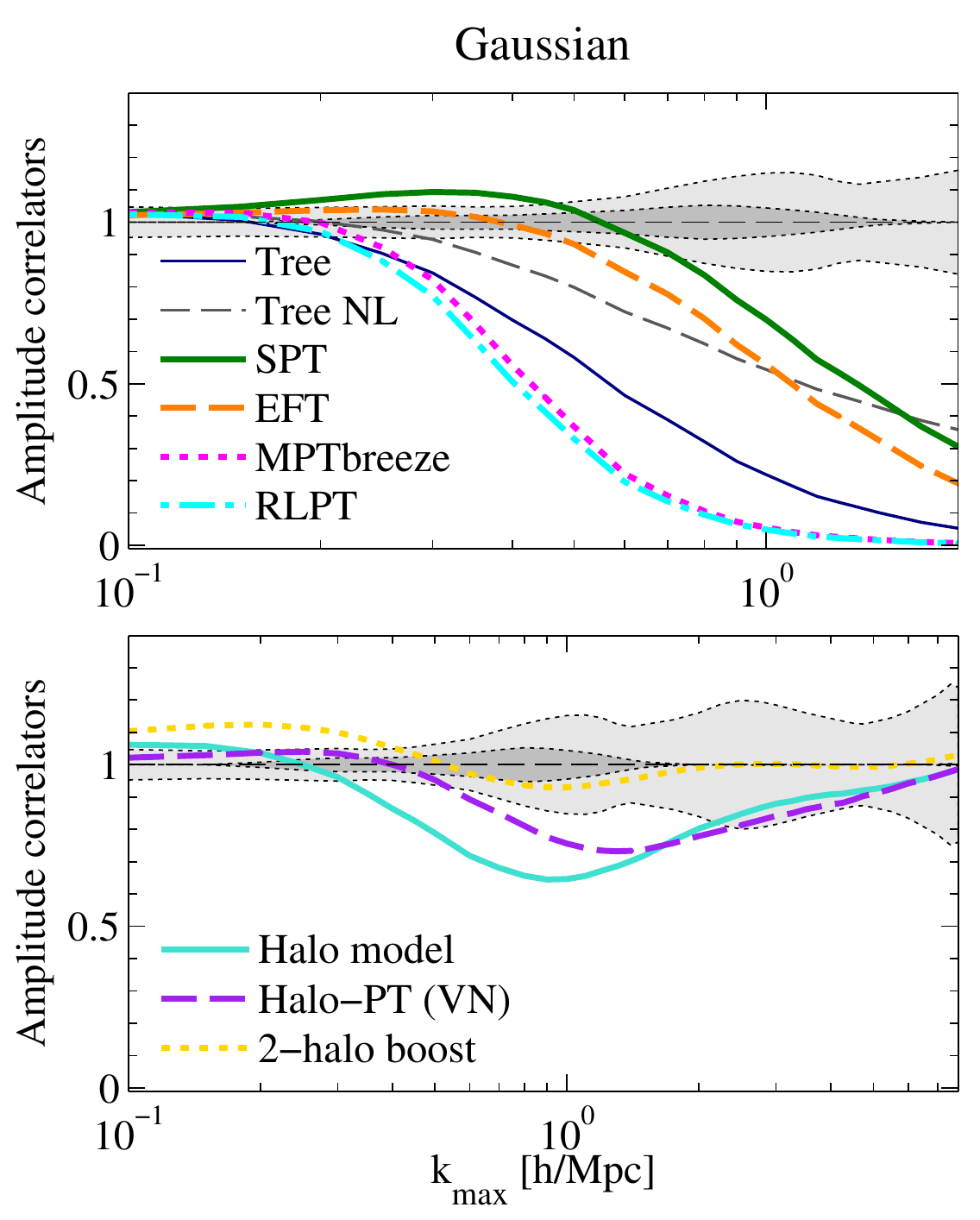}  
\caption{Amplitude correlators at $z=1$ between the benchmark ``three-shape'' model and all perturbative approaches (upper) and halo models (lower). Shape correlators are higher.}
\label{combkmaxpt}
\end{center}
\end{figure}

\begin{figure}
\begin{center}
\includegraphics[width=\linewidth]{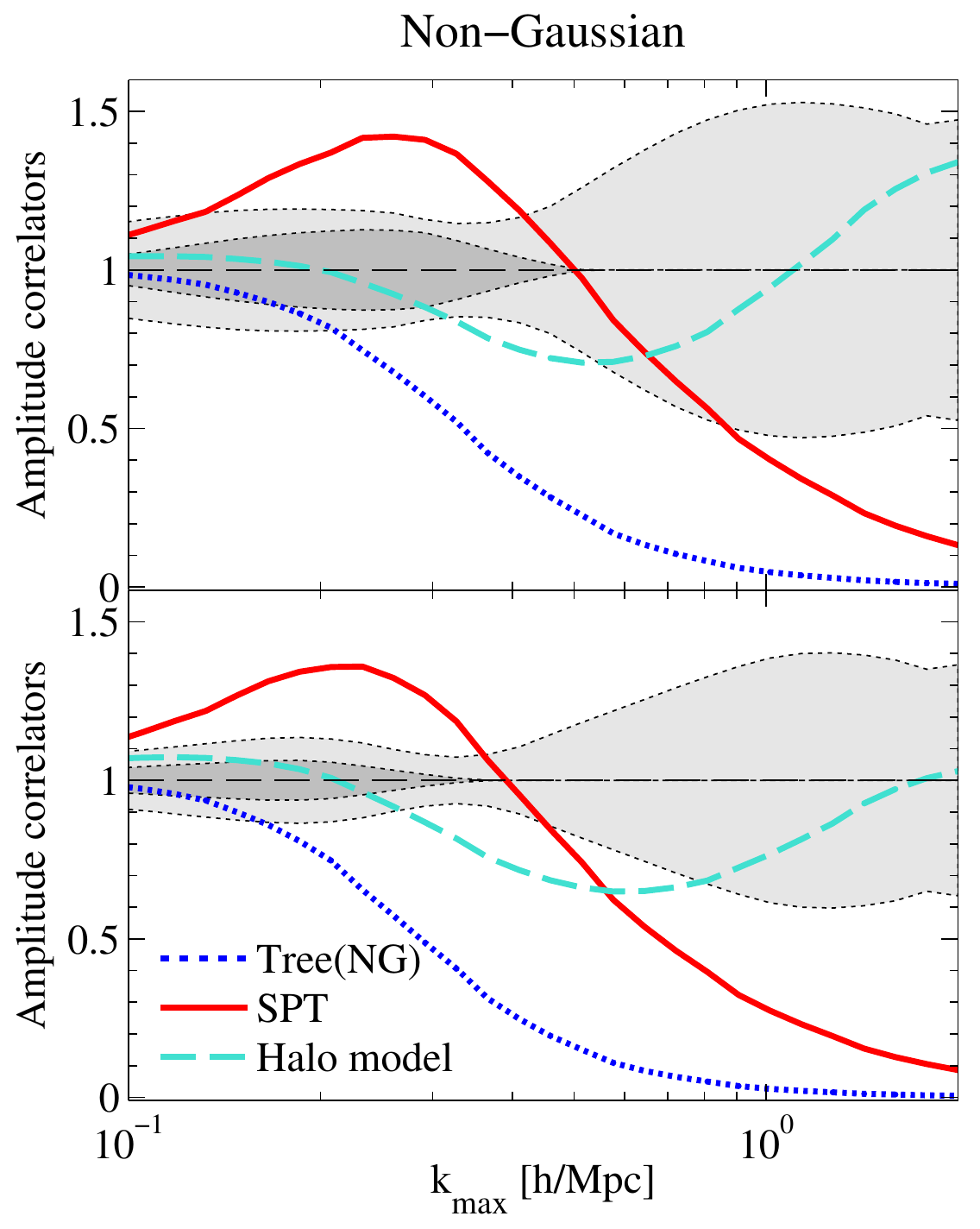}  
\caption{Amplitude correlators at $z=1$ between the benchmark ``three-shape'' model for $\Delta B_{\text{NG}}^{\text{local}}$ (top) and $\Delta B_{\text{NG}}^{\text{eq}}$ (bottom) respectively, and the tree level, one loop SPT and halo model non-Gaussian corrections. Shape correlators are higher.}
\label{combkmaxloc}
\end{center}
\end{figure}

\smallskip
\section{VII. Testing theoretical approaches}
We next use the calibrated ``three-shape'' model [Eq.~(\ref{shapes})]  as a reference benchmark against which to test the accuracy of other perturbative and halo models using the amplitude $\mathcal{A}$ and shape correlators $\mathcal{S}$. We choose to plot our results in terms of the cumulative version of the correlators, because we are interested in determining up to which scales each model can be trusted. As the bispectrum is smooth, and both the theoretical models and simulations can be described by simple shape functions on slices of constant $k_1+k_2+k_3$ (Ref.~\cite{bislong}), we are confident that this method provides an adequate comparison between models and simulations. The results at $z\,$=$\,1$ are shown in Fig.~\ref{combkmaxpt} for the Gaussian simulations with light and dark grey areas representing, respectively, uncertainties between simulation realisations and the 
inaccuracy in $B_{\text{3-shape}}$  (see Ref.~\cite{bislong} for further details).   For the perturbative theories in Fig.~\ref{combkmaxpt} (upper panel), we see the generic result that SPT produces an excess on mildly nonlinear scales before starting to decay in the nonlinear regime.   This excess is removed in EFT by adding a counterterm calibrated to simulations, thus offering reasonable accuracy over an extended $k$-range. RPT (\textsc{MPTbreeze}) and RLPT produce accurate results on very large scales (low $k$), with the accuracy increasing as the number of loops is increased. (We also calculated the two-loop bispectrum in the \textsc{MPTbreeze} formalism for a few slices). Both methods cut off the excess seen in SPT with physically-motivated exponential functions, which causes the sharp decay beyond the linear regime.  Of all these methods, we conclude that EFT succeeds in extending the validity range furthest into the nonlinear regime.   However, the much simpler nonlinear tree-level  approximation (``treeNL'') also provides a useful extrapolation for the bispectrum at higher $k$. 
More quantitatively, we find that at $z=1$ SPT is accurate at the 5\% level up to $k = 0.14 \, h/$Mpc; EFT extends this to $k = 0.36 \, h/$Mpc, while \textsc{MPTbreeze} and RLPT are limited to $k = 0.21 \, h/$Mpc, $k = 0.19 \, h/$Mpc respectively, comparable with the treeNL approach ($k = 0.22 \, h/$Mpc).

Although the standard halo model provides an adequate fit for the bispectrum at $z=0$, at higher redshifts a significant deficit appears on intermediate scales, as shown at $z\,$$=$$\,1$  in Fig.~\ref{combkmaxpt} (lower panel).   For $z \ge 2$, this deficit becomes larger than a factor of 2, reflecting the incorrect underlying growth rate of the two-halo term, a problem shared by the halo-PT model, which is also plotted.   Boosting the two-halo signal at higher redshift can alleviate the intermediate deficit, but it creates an  excess at low $k$.   The standard halo model clearly requires a substantially modified two-halo term, combined with an improved perturbative approximation (such as the nonlinear tree-level or the EFT models), any of which can be easily incorporated into the ``three-shape'' model. 

We extend these results for the local and equilateral types of primordial non-Gaussianity using Eq.~(\ref{shapesNG}), with the results plotted in Fig.~\ref{combkmaxloc}. For both types of primordial non-Gaussianity considered, we show that we obtain a similar conclusion as for the Gaussian simulations --- the one-loop SPT exhibits an excess of power before quickly decaying, while the halo model has a deficit on intermediate scales at high redshifts, becoming accurate again on small scales.

\section{VIII. Conclusion}
In this paper we have weighed the relative merits of a broad range of theoretical bispectrum models through a quantitative comparison with the full 3D matter bispectrum estimated from $N$-body simulations. Inspired by the three canonical shapes we identified in the halo model, we have developed a simple phenomenological ``three-shape'' model which can achieve a good match to the  $N$-body bispectrum for all redshifts $z<3$. We have shown that this simple model can be easily extended to model primordial non-Gaussianity by identifying the same shapes in the non-Gaussian part of the matter bispectrum on the same scales as in the Gaussian bispectrum. Thus, we have performed a 3D comparison between theoretical models and $N$-body simulations for primordial non-Gaussianity.
In the future, we will improve the precision of the ``three-shape'' model by calibrating it on higher-resolution simulations: this will be useful to test parameter dependencies and eventually to provide a \textsc{Halofit}-like bispectrum model, with applications e.g. to weak gravitational lensing. We will extend this modelling to the bispectrum of biased tracers, as well as further types of primordial non-Gaussianity.

\section{Acknowledgements}
We wish to thank James Fergusson and Donough Regan for useful discussions.
T.G. acknowledges support from the Kavli Foundation and STFC Grant No. ST/L000636/1.
This work used the  COSMOS SMP supercomputer at DAMTP,  Cambridge, operated
on behalf of the STFC DiRAC HPC Facility, and it was supported by STFC Grants No. ST/J005673/1, No. ST/H008586/1, No.
ST/K00333X/1 and No. STM007065/1.

\bibliography{Bib}{}

\end{document}